\documentclass[prB,twocolumn,amsmath,amssymb,superscriptaddress,aps]{revtex4-1}
\usepackage{amssymb} 
\usepackage{amsmath}
\usepackage{mathtools}
 
\begin{document}

\title{Magnetic anisotropy and exchange interactions of two-dimensional FePS$_3$, NiPS$_3$ and MnPS$_3$ from first principles calculations} 

\author{Thomas Olsen}
\affiliation{Computational Atomic-scale Materials Design (CAMD), Department of Physics, Technical University of Denmark, DK-2800 Kgs. Lyngby, Denmark}
\email{tolsen@fysik.dtu.dk}

\date{\today}

\begin{abstract}
The van der Waals bonded transition metal phosphorous trichalcogenides FePS$_3$, NiPS$_3$ and MnPS$_3$ have recently attracted renewed attention due to the possibility of exfoliating them into their monolayers. Although the three compounds have similar electronic structure, the magnetic structure differs due to subtle differences in exchange and magnetic anisotropy and the materials thus comprise a unique playground for studying different aspects of magnetism in 2D. Here we calculate the exchange and anisotropy parameters of the three materials from first principles paying special attention to the choice of Hubbard parameter U. We find a strong dependence of the choice of U and show that the calculated Néel temperature of FePS$_3$ varies by an order of magnitude over commonly applied values of U for the Fe $d$-orbitals. The results are compared with parameters fitted to experimental spin-wave spectra of the bulk materials and we find excellent agreement between the exchange constants when a proper value of U is chosen. However, the anisotropy parameters are severely underestimated by DFT and we discuss possible origins of this discrepancy.
\end{abstract}

\maketitle

\section{Introduction}
The discovery of ferromagnetic order in two-dimensional (2D) CrI$_3$ \cite{Huang2017a} in 2017 has initiated a vast interest in the field of 2D magnetism \cite{Burch2018, Soriano2020, Sethulakshmi2019, Gibertini2019}. Subsequently, several other magnetic van der Waals bonded compounds \cite{McGuire2017a} have been exfoliated to the monolayer limit and shown to exhibit 2D magnetic order \cite{Burch2018,Bonilla2018,Fei2018,Pedersen2018}. It is, however, not obvious that monolayers exfoliated from magnetic van der Waals bonded materials retain the magnetic order in general. The reason is that 2D materials cannot exhibit a spontaneously broken spin-rotational symmetry \cite{MerminWagner} and either (weak) interlayer interactions or magnetic anisotropy are thus vital ingredients for magnetic order in van der Waals bonded materials. Typically, only the latter case will result in magnetic order for the isolated monolayer and spin-orbit interactions (which are responsible for magnetic anisotropy) thus comprise a crucial prerequisite for 2D magnetism. 

The transition metal thiophosphates MPS$_3$ (M=Fe,Ni,Mn) comprise a particular interesting class of van der Waals magnets \cite{Jernberg1984,Joy1992,Wildes1998,Wildes2015,Lancon2016,Lancon2018,Xing2019,Kang2020} that exhibit rather distinct magnetic properties in the monolayer limit. In bulk form they all exhibit anti-ferromagnetic order in the individual planes with Néel temperatures of 123 K, 155 K and 78 K for FePS$_3$, NiPS$_3$ and MnPS$_3$ respectively \cite{Joy1992}. However only FePS$_3$ has been demonstrated to retain its magnetic order in the case monolayers with the Néel temperature being reduced to 104-118 K \cite{Wang2016h, Lee2016}. This can be understood from the fact that bulk FePS$_3$, exhibits a strong out-of-plane easy-axis \cite{Joy1992}, which breaks the rotational symmetry and allows for magnetic order in the monolayer limit. In contrast, magnetic order in NiPS$_3$ has been shown to persist in bilayers, but disappears for a monolayer \cite{Kim2019}. This is expected from the fact that bulk NiPS$_3$ exhibits an easy-plane coinciding with the atomic layers and if the anisotropy is maintained in the monolayer limit there is a residual rotational symmetry, which deteriorates magnetic order as a consequence of the Mermin-Wagner theorem. Finally, bulk MnPS$_3$ exhibits an out-of-plane easy axis and would be expected to exhibit magnetic order in the monolayer limit \cite{Wildes1998}. However, to our knowledge there are no reports on the magnetic order (or its absence) in monolayers of MnPS$_3$ although magnetic order has been demonstrated in bilayers \cite{Kim2019a}.

From the computational community there has been
a vivid search for new 2D magnets based on high throughput first principles calculations \cite{Mounet2018,Miyazato2018,Haastrup2018,Torelli2019c,Torelli2020a,Botana2019,Kabiraj2020} with various attempts of predicting magnetic critical temperatures for magnetic order. Such computations do, however, rely crucially on the accuracy of the applied method. In particular, for methods based on density functional theory (DFT) different choices of exchange-correlation functional may lead to predicted exchange constants that differ by a factor of three \cite{olsen_mrs}. Moreover, for 2D materials it is crucial to obtain accurate predictions for the magnetic anisotropy, which plays a prominent role in the theory of magnetic order. In this paper we address the accuracy of first principles calculations for exchange parameters and anisotropy constants. The calculations are performed on the three 2D compounds FePS$_3$, NiPS$_3$ and MnPS$_3$, since these materials provide convenient examples of different types of magnetic order and comprise realizations of easy-axis magnetization and easy-plane magnetization. We pay particular attention to the effect of the value of U used in DFT+U calculations and show that different choices can lead to significantly different predictions for the magnetic parameters. Finally, we show that a Heisenberg model including single-ion anisotropy and anisotropic exchange is not able to reproduce the large spin-wave gaps observed for the bulk compounds.

The paper is organized as follows. In Sec. \ref{sec:theory} we provide the basic theoretical framework that allow us to determine exchange and anisotropy parameters from DFT calculations. In Sec. \ref{sec:computational} we summarize the computational details of the calculations and in \ref{sec:results} we provide the results. Sec. \ref{sec:discussion} provides a summary and a discussion of the results.

\section{Theory}\label{sec:theory}
Whereas DFT can usually faithfully predict the magnetic ground state of a given material, the thermodynamical properties are largely inaccessible by direct computations. Instead one is led to define a magnetic model that captures the essential interactions and is simple enough to allow for thermodynamical predictions. For insulators the Heisenberg model \cite{Yosida1996} has proven highly successful in providing qualitative predictions for phase transitions and if the model parameters are determined by DFT the model acquires quantitative predictive power \cite{Schmitt2014,Xiang2013,Olsen2017,Torelli2019}. For the purpose of investigating critical temperatures in 2D materials we thus consider the model Hamiltonian
\begin{align} \label{eq:H}
H=-\frac{1}{2}\sum_{ij}J_{ij}\mathbf{S}_i\cdot\mathbf{S}_j-\frac{1}{2}\sum_{ij}\lambda_{ij}S_i^zS_j^z-A\sum_i(S_i^z)^2,
\end{align}
where the sums run over magnetic atoms in the compound and $\mathbf{S}_i$ is the spin operator for site $i$. $J_{ij}$ denotes the isotropic exchange between site $i$ and $j$, $\lambda_{ij}$ is the anisotropic exchange and $A$ denotes the strength of single-ion anisotropy. For 2D materials it is vital to include the anisotropy terms due to the Mermin-Wagner theorem. Since the anisotropic exchange and single-ion anisotropy only involve the $z$-component of spin operators we have implicitly assumed magnetic isotropy in the $xy$-plane, which is taken to coincide with the atomic plane. We have neglected off-diagonal exchange terms (for example terms proportional to $S_i^xS_j^y$). Such terms may give rise to interesting physical effects such as chiral magnetic interactions and Kitaev terms in the Hamiltonian, but will not be considered here since we expect that these have minor influence on the critical temperature.

The Heisenberg model \eqref{eq:H} can be analyzed, for example, from renormalized spinwave theory \cite{Yosida1996,Yasuda2005,Gong2017b, Lado2017} or classical Monte Carlo simulations \cite{Akturk2017, Torelli2019, Lu2019}. The former case comprises a full quantum mechanical treatment that is accurate at low temperatures. However, spinwave interactions are treated at the mean-field level and may become inaccurate in the vicinity of the critical temperature where the number of spin-waves increases dramatically \cite{Yasuda2005}. In contrast, classical Monte Carlo simulations completely neglects quantum effects, but includes all correlation in the model. At elevated temperatures (close to the critical temperature in particular) quantum effects tend to be quenched and the classical analysis is expected to become accurate \cite{Torelli2019} - perhaps with the exception of spin-$1/2$ materials \cite{Yasuda2005}. In the present work we have thus applied classical Monte Carlo simulations to extract critical temperatures.

In order make quantitative predictions for real materials the parameters in the model \eqref{eq:H} need to be determined from first principles calculations. Including $n$-nearest neighbor couplings in the model yields $2n+1$ parameters that can be determined from $2n+2$ DFT calculations involving different spin configurations. Since the anisotropy parameters arise from spin-orbit coupling one may can consider $n+1$ spin configurations without spin-orbit coupling and then obtain $2n+2$ total energies with non-selfconsistent spin-orbit coupling by orienting the exchange-correlation magnetic field parallel and orthogonal to the atomic plane for each configuration. 

\begin{figure}[t]
	\includegraphics[width = 0.45\textwidth]{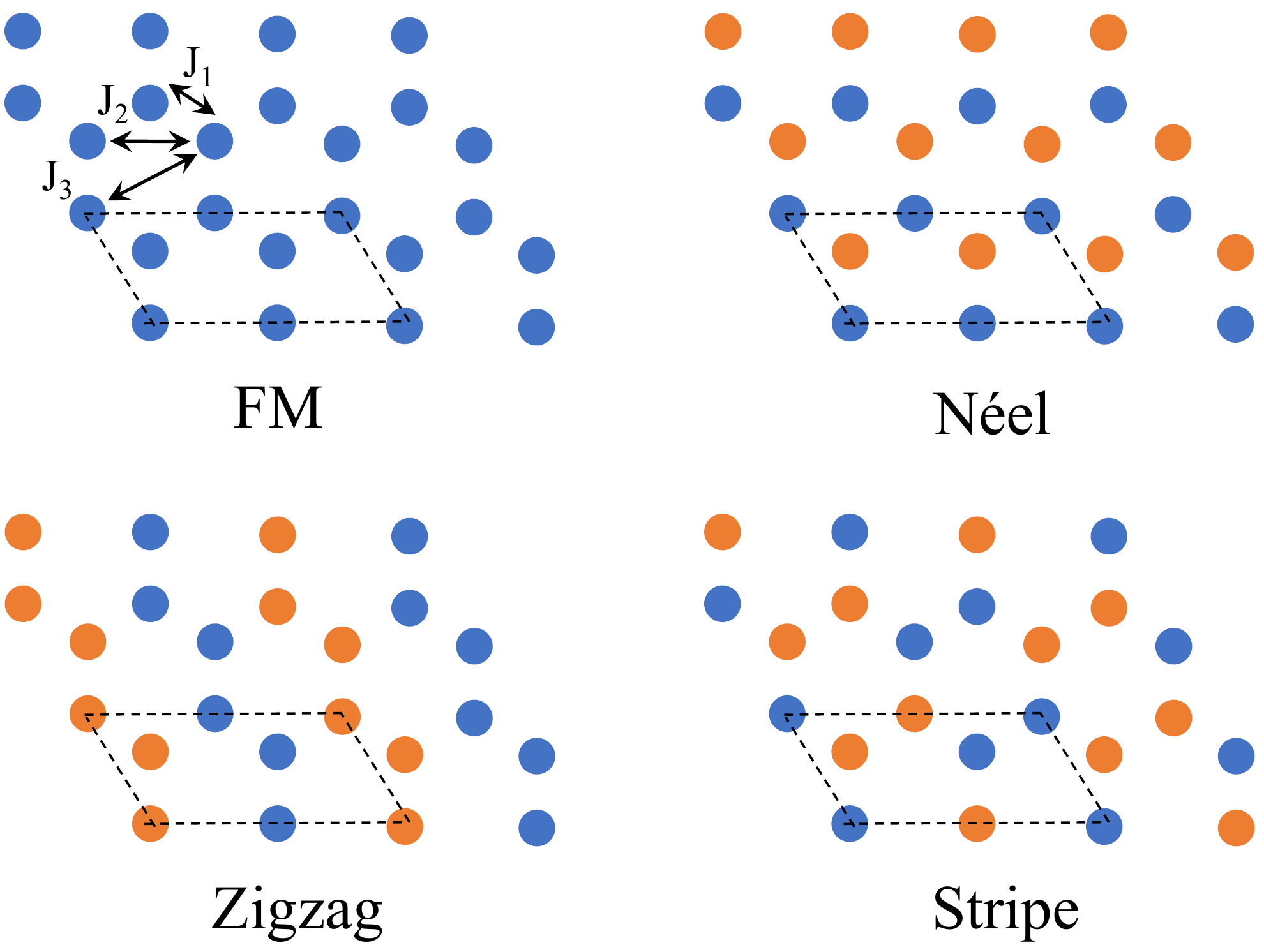}
	\caption{Spin configurations used for the energy mapping analysis. The different colors denote different states of the atomic magnetic moments. The unit cell in each case is shown by dashed lines and contains 4 magnetic transition metal atoms.}
	\label{fig:spin_conf}
\end{figure}

The third nearest neighbor exchange coupling has previously been shown to be particular important for the transition metal phosphorous trichalcogenides and we thus consider all interactions up to third nearest neighbor in the model \eqref{eq:H}. It should be noted that there are 3 nearest and third nearest neighbors while there are 6 second nearest neighbors. In contrast to previous works we also calculate the anisotropy parameters, which are crucial for obtaining reliable estimates of the critical temperature. We thus consider the four spin configurations shown in Fig. \ref{fig:spin_conf}, which are used to extract the exchange parameters as
\begin{align}
J_1 =& \frac{1}{4S^2}(E_\mathrm{Stripy}^\parallel - E_\mathrm{FM}^\parallel + E_\mathrm{Neel}^\parallel - E_\mathrm{Zigzag}^\parallel)\\
J_2 =& \frac{1}{8S^2}(E_\mathrm{Stripy}^\parallel - E_\mathrm{FM}^\parallel - E_\mathrm{Neel}^\parallel + E_\mathrm{Zigzag}^\parallel)\\
J_3 =& \frac{1}{3S^2}(E_\mathrm{FM}^\parallel - E_\mathrm{Neel}^\parallel)-J_1\\
\lambda_1 =& \frac{1}{4S^2}(\Delta E_\mathrm{Stripy} - \Delta E_\mathrm{FM} + \Delta E_\mathrm{eel} - \Delta E_\mathrm{Zigzag})\\
\lambda_2 =& \frac{1}{8S^2}(\Delta E_\mathrm{Stripy} - \Delta E_\mathrm{FM} - \Delta E_\mathrm{Neel} + \Delta E_\mathrm{Zigzag})\\
\lambda_3 =& \frac{1}{3S^2}(\Delta E_\mathrm{FM} - \Delta E_\mathrm{Neel})-\lambda_1\\
A =& \lambda_2 - \frac{1}{2S^2}(\Delta E_\mathrm{Stripy} + \Delta E_\mathrm{Zigzag}),
\end{align}
where $E_\alpha^\parallel$ is the total energy per magnetic atom of configuration $\alpha$ with the exchange-correlation magnetic field aligned in the atomic plane. $\Delta E_\alpha=E_\alpha^\perp-E_\alpha^\parallel$ is the energy difference per magnetic atom for spin state $\alpha$ between spins aligned in the plane ($E^\parallel_\alpha)$) and spins aligned out of plane ($E^\perp_\alpha)$). We note that these parameters were extracted by mapping total energies to the {\it classical} Heisenberg model. It has previously been shown that for nearest neighbor exchange only it is possible to map the total energies directly to the quantum mechanical Heisenberg model, which yields exchange couplings that are 3-7 {\%} lower than those obtained from the classical model \cite{Torelli2020}. However, in the present case a full quantum mechanical energy mapping analysis would be non-trivial and we will stick with the classical parameters stated above in the following.

\section{Computational details}\label{sec:computational}
All DFT calculations were obtained with the electronic structure code GPAW using the projector-augmented wave method and a plane wave basis \cite{Enkovaara2010a,Larsen2017}. Spin-orbit coupling was included non-selfconsistently \cite{Olsen2016a} and a direction for the spins was chosen by rotating the spin-dependent mean-field along the desired direction. We used the PBE+U functional and a plane wave cutoff of 600 eV. The unit cell in all calculations were chosen as shown in Fig. \ref{fig:spin_conf} and the Brillouin zone sampling was done on a $\Gamma$-centered 6x12 grid. In calculations with PBE+U we put the value of U on the transition metal $d$-orbitals. The structures were relaxed until all forces are below 0.05 eV/Å.

In order to obtain critical temperatures of FePS$_3$ we have performed classical Monte Carlo simulations using the Metropolis algorithm with a $20\times20$ repetition of the minimal unit cell containing two magnetic sites and periodic boundary conditions. We used 100,000 Monte Carlo steps, where each step involves a random spin flip of all sites in the lattice and the total energy was extracted from an average over the last 20,000 steps. The heat capacity was then evaluated by finite difference between the energies at neighboring temperatures and the the critical temperature extracted from a Lorentzian fit in the vicinity of the maximum of the heat capacity.

\section{Results}\label{sec:results}

\subsection{Magnetic ground state}
\begin{figure*}[t]
	\includegraphics[width = 0.8\textwidth]{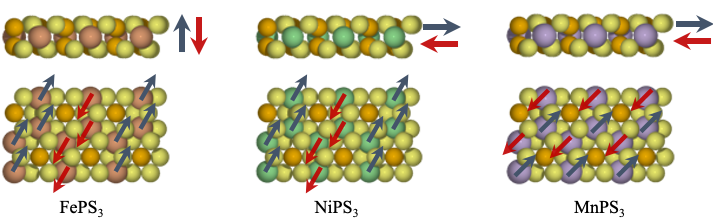}
	\caption{Magnetic ground state of FePS$_3$, NiPS$_3$ and MnPS$_3$. FePS$_3$ and NiP$_3$ exhibits a zigzag type anti-ferromagnetic order whereas MnPS$_3$ has Néel type anti-ferromagnetic order. Only FePS$_3$ has an easy axis orthogonal to the atomic plane.}
	\label{fig:structure}
\end{figure*}
The magnetic ground states of FePS$_3$, NiPS$_3$ and MnPS$_3$ are shown in Fig. \ref{fig:structure}. All of the compounds are anti-ferromagnetic, but only MnPS$_3$ acquires the Néel state where each magnetic site is anti-aligned with all nearest neighbors. In contrast, the ground states of FePS$_3$ and NiPS$_3$ exhibit Zigzag-type ordering (see Fig. \ref{fig:spin_conf}), where each transition metal atom is aligned with two nearest neighbors and anti-aligned with one nearest neighbor, which indicates ferromagnetic nearest neighbor exchange. We note that only the Ferromagnetic and Néel configurations can be represented in the primitive (non-magnetic) unit cell of the lattice.% and we have performed a full relaxation of the four magnetic structures in the unit cells shown in Fig. \ref{fig:structure}. 
%In general, however, the relaxed atomic positions are rather insensitive to the magnetic state enforced in the relaxation. 
The magnetic ground state in the three compounds are insensitive to the choice of U in a DFT+U treatment when U is chosen up to 7 eV. However, as will be shown below the magnitude of the magnetic interactions depend strongly on U.

The relative values of exchange coupling constants in the classical Heisenberg model \eqref{eq:H} can be related to the magnetic ground state. For example, if one neglects the contributions from anisotropy the Néel state will be favored over the FM state if $J_1+J_3<0$, the Zigzag state is favored over the FM state if $J_1+4J_2+3J_3<0$ and the Striped state is favored over the FM state if $J_1+2J_2<0$. Moreover, the Zigzag state will be favored over the Néel state if $J_1>2J_2$, which is the case for FePS$_3$ and NiPS$_3$ as shown below.

\begin{figure*}[t]
	\includegraphics[width = 0.32\textwidth]{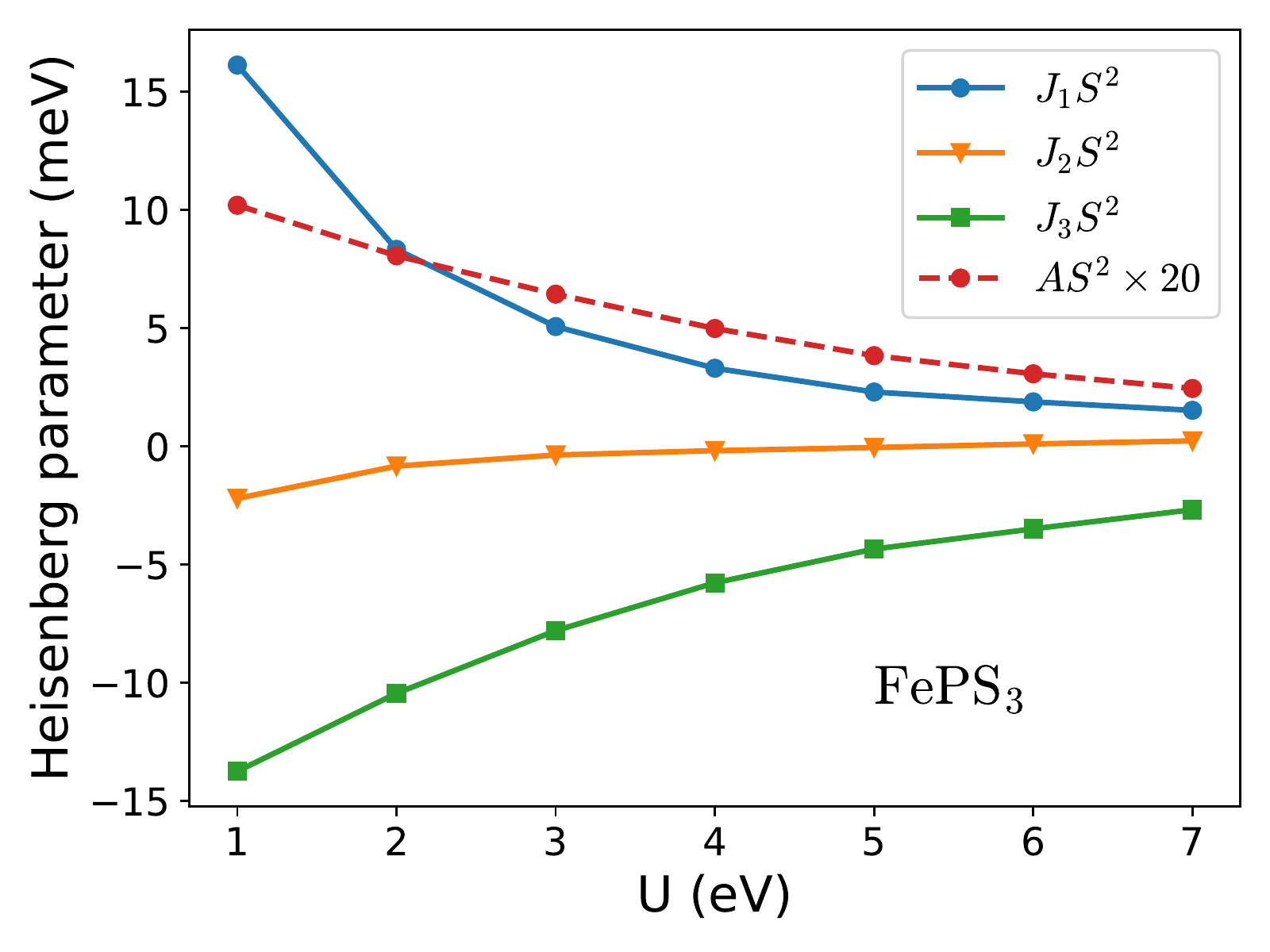}
	\includegraphics[width = 0.32\textwidth]{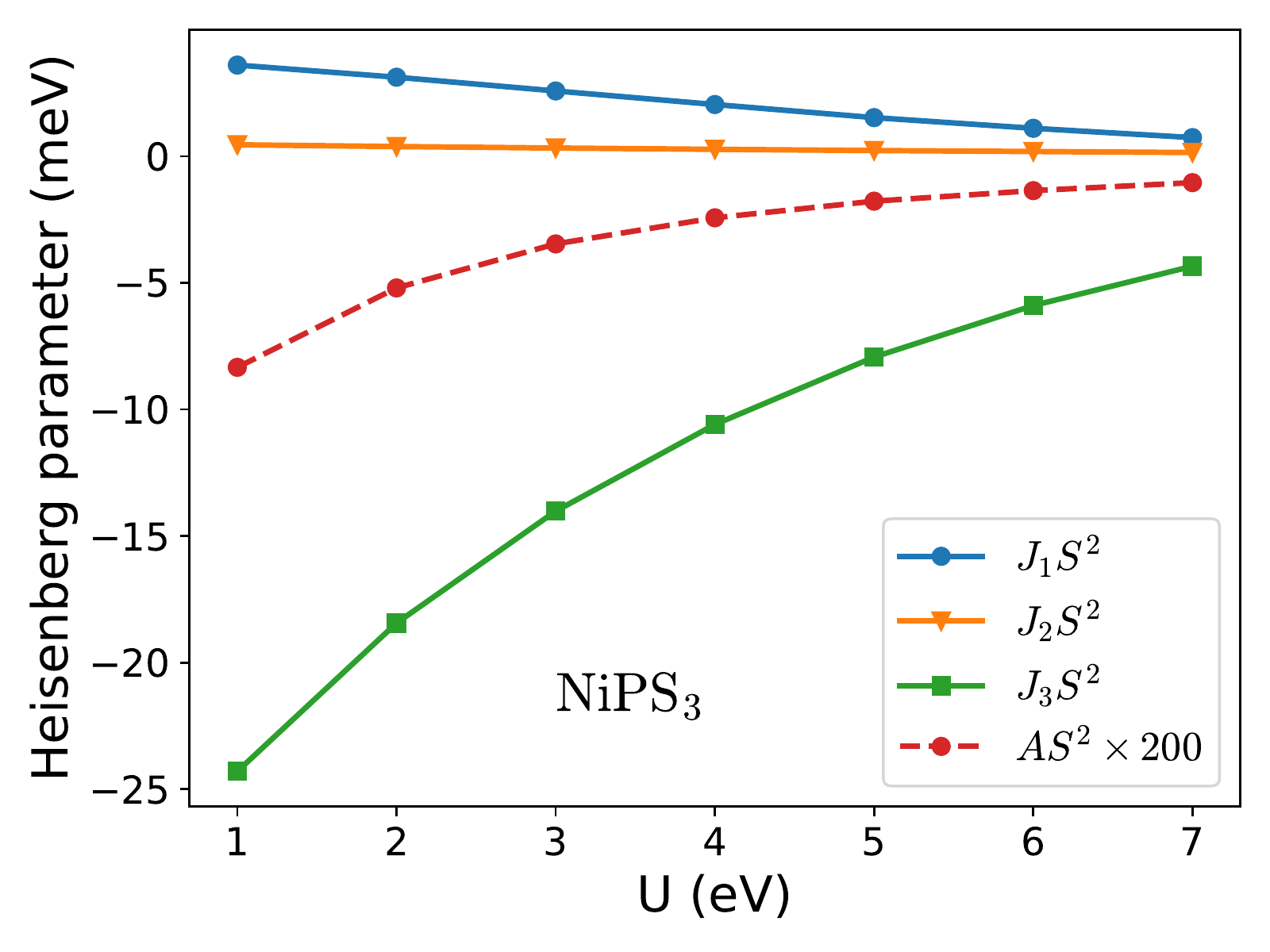}
	\includegraphics[width = 0.32\textwidth]{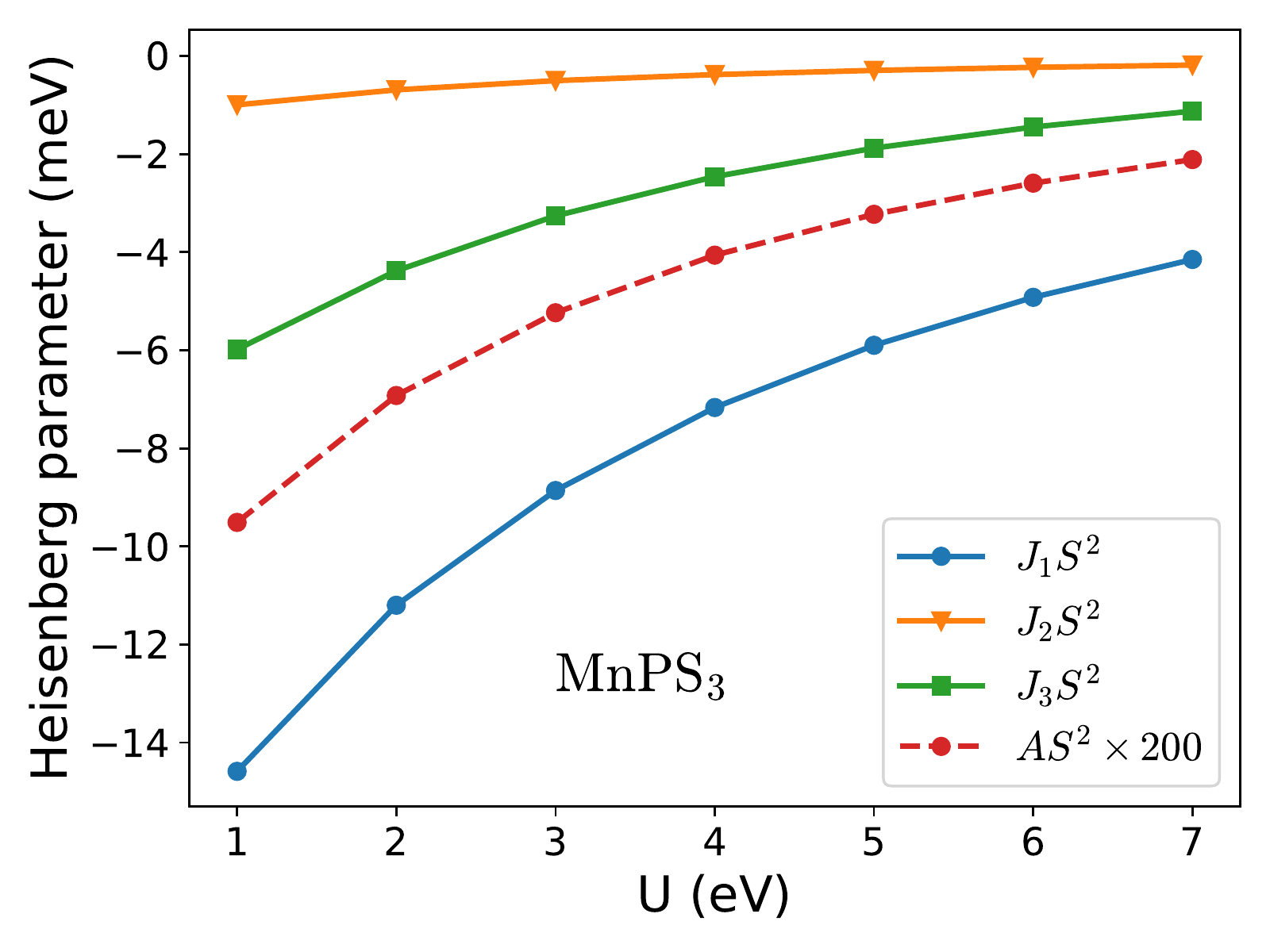}
	\caption{Calculated isotropic exchange parameters and single-ion anisotropy as a function of U for FePS$_3$, NiPS$_3$, and MnPS$_3$.}
	\label{fig:exchange}
\end{figure*}
Due to the Mermin-Wagner theorem, a magnetic easy-axis is required for magnetic order at finite temperatures. In the present case only FePS$_3$ is predicted to have an easy axis, whereas NiPS$_3$ and MnPS$_3$ both have easy planes coinciding with the atomic planes. Monolayers of FePS$_3$ have indeed been found to exhibit anti-ferromagnetic order up to 118 K in experiments \cite{Lee2016}, whereas the magnetic order in NiPS$_3$ has been shown to be quenched in the monolayer limit down to 10 K \cite{Kim2019}. Bulk MnPS$_3$ has been argued to be largely isotropic \cite{Joy1992}, which is expected due to the orbitally closed $d$-shell with $S=5/2$. We find an anisotropy energy of 0.053 meV per Mn atom (energy difference between spins oriented in-plane and out-of-plane). This is, however, slightly larger than the value of 0.037 meV found for the $S=1$ material NiPS$_3$, which indicates that {\it a priori}  prediction of spin-orbit effects is highly challenging. In addition, both values are an order of magnitude smaller than the value of FePS$_3$, which is found to be 0.45 meV per Fe atom.

\subsection{Heisenberg parameters and critical temperatures}
\begin{table*}[tb]
  \begin{center}
\begin{tabular}{l|ccccccc}
  Material  & $J_1$ & $J_2$ & $J_3$ & $\lambda_1$ & $\lambda_3$ & $\lambda_3$ & A \\
    \hline
    \hline
FePS$_3$ (U=2 eV) & 2.1 & -0.21  & -2.6  & -4.1$\times10^{-3}$ & 1.1$\times10^{-3}$ & 2.5$\times10^{-3}$ & 0.101  \\
FePS$_3$ (experimental) \cite{Lancon2016} & 2.92 & -0.08  & -1.92  & - & - & - & 2.66 \\   
NiPS$_3$ (U=3 eV) & 2.6 &  0.32  & -14 & -0.32$\times10^{-3}$ & -0.51$\times10^{-3}$ & -0.25$\times10^{-3}$ & -0.018  \\
NiPS$_3$ (experimental) \cite{Lancon2018} & 3.8 &  -0.2  & -13.8 & - & - & - & 0.3  \\
MnPS$_3$ (U=3 eV) & -1.42 & -0.081  & -0.52 & -1.2$\times10^{-3}$ & -0.19$\times10^{-3}$ & 0.37$\times10^{-3}$ & -0.0035   \\
%MnPS$_3$ (U=5 eV, a=6.15 Å) & -2.3 & -0.160  & -0.96 & 0.141$\times10^{-3}$ & -0.25$\times10^{-3}$ & -0.109$\times10^{-3}$ & -0.0076  \\
%MnPS$_3$ (U=5 eV, a=6.15 Å) & -0.94 & -0.044  & -0.30 & 1.90$\times10^{-3}$ & -0.068$\times10^{-3}$ & 0.27$\times10^{-3}$ & -0.0027  \\
%MnPS$_3$ (U=5 eV, a=5.88 Å) & -1.60 & -0.076  & -0.36 & -0.0047$\times10^{-3}$ & -0.099$\times10^{-3}$ & -0.0170$\times10^{-3}$ & -0.0019\\
MnPS$_3$ (experimental) \cite{Wildes1998} & -1.54 & -0.14  & -0.36 & - & - & - & 0.0086
\end{tabular}
\end{center}
\caption{Calculated Heisenberg parameters at the value of U that provides the best match to experimental parameters. Experimental values were obtained by fitting to spin-wave spectra of the bulk 3D materials and are taken from Ref.  \cite{Lancon2018}.}
\label{tab:heisenberg}
\end{table*}
The Heisenberg parameters of Eq. \eqref{eq:H} has been calculated for the three materials studied in this work. However, the parameters turn out to be rather sensitive to the value of U used in a DFT+U approach. In Fig. \ref{fig:exchange} we show the isotropic exchange constants as well as the single-ion anisotropy as a function of U for the three compounds. In all cases we observe a reduction of the exchange constants by a factor of 2-4 when increasing U from 1 eV to 5 eV. The effect is most dramatic in FePS$_3$ where $J_1S^2$ decreases from 16 meV to 2.3 meV. To rationalize this trend one may argue that larger values of U tends to increase orbital localization and therefore decrease the overlap between wavefunctions. In the case of direct exchange interactions this will in general decrease exchange integrals and therefore decrease the magnitude of exchange interactions. For superexchange the exchange constants are roughly given by $-t^2/U$ where $t$ is a hopping matrix element. In that case one would also expect a decreased magnitude of exchange coupling constants. In reality the effect of U may, however, be significantly more complicated than this simplistic picture and in CrI$_3$, for example, it has been shown that increasing $U$ tends to increase the magnitude of exchange coupling constants \cite{Torelli2019c}. Nevertheless, for the case of FePS$_3$, NiPS$_3$ and MnPS$_3$ we observe a sizeable decrease in exchange coupling constants when increasing the Hubbard parameter. In addition we also observe a significant decrease in single-ion anisotropy with increasing Hubbard corrections. This can be rationalized from the fact that the spin-orbit coupling is completely dominated by spherical contributions to the crystal field in the vicinity of the nuclei. The magnetic anisotropy thus arises from hybridization effects, which are suppressed by the Hubbard corrections. A similar picture was observed for CrI$_3$ although in that case the anisotropic exchange increases with increasing Hubbard corrections resulting in an overall increase in the magnetic anisotropy \cite{Torelli2019c}.  The predicted values of the anisotropic exchange constants are neglectable in all three  cases.

In Tab. \ref{tab:heisenberg} we present the parameters calculated with PBE+U using values of U that provides the best agreement with experimentally determined values for the bulk material (also displayed for reference). For all three materials we note that the magnitude of the second-nearest neighbor coupling $J_2$ is much smaller than $J_1$ and $J_3$. The magnitudes of $J_1$ and $J_3$ in FePS$_3$ are similar and the fact that $J_1>0$ and $J_3<0$ determines the Zigzag state as the magnetic ground state. In the case of of NiPS$_3$ the anti-ferromagnetic $J_3$ is completely dominating and the positive $J_1$ again determines the ground state to have Zigzag order. In contrast, MnPS$_3$ shows all anti-ferromagnetic exchange coupling constants and exhibits Néel-type order. The results are in reasonable agreement with experimental values at the chosen values of the Hubbard parameter $U$, but will deviate significantly if other values are applied. We note that the results appear to be in disagreement with previous PBE+U calculations for MnPS$_3$ \cite{Sivadas2015} that yielded $J_1=-1.58$ meV, $J_2=-0.08$ meV and $J_3=-0.46$ meV (note the different convention for $J_i$ in Ref. \cite{Sivadas2015}) using a Hubbard parameter of 5 eV. In that work however, the experimental lattice parameter of $a=5.88\;\AA$ was used whereas we have used the PBE relaxed structure with $a=6.15\;\AA$. If we base the calculations on a relaxed structure using the experimental lattice parameter and U = 5 eV we obtain $J_1=-1.6$ meV, $J_2=-0.076$ meV and $J_3=-0.36$ meV, which is in very good agreement with the experimental values as well as the previous theoretical predictions \cite{Sivadas2015}. Redoing the calculations with experimental lattice parameter and U = 3 eV, however, lead to parameters that are roughly twice the experimental values. The exchange parameters are thus highly sensitive to correct lattice parameter and resulting interatomic distances although a modified value of $U$ can be applied to correct for the error originating from an overestimated lattice parameter. In the respect, the current example of MnPS$_3$ is a rather extreme example where PBE overestimates the lattice parameter by 4.6 {\%}. In any case, the results are seen to depend strongly on the choice of U and the agreement with experimental values thus appears fortuitous, since it is vital to choose the correct value of U, which is not known {\it a priori}. 

\begin{figure}[t]
	\includegraphics[width = 0.45\textwidth]{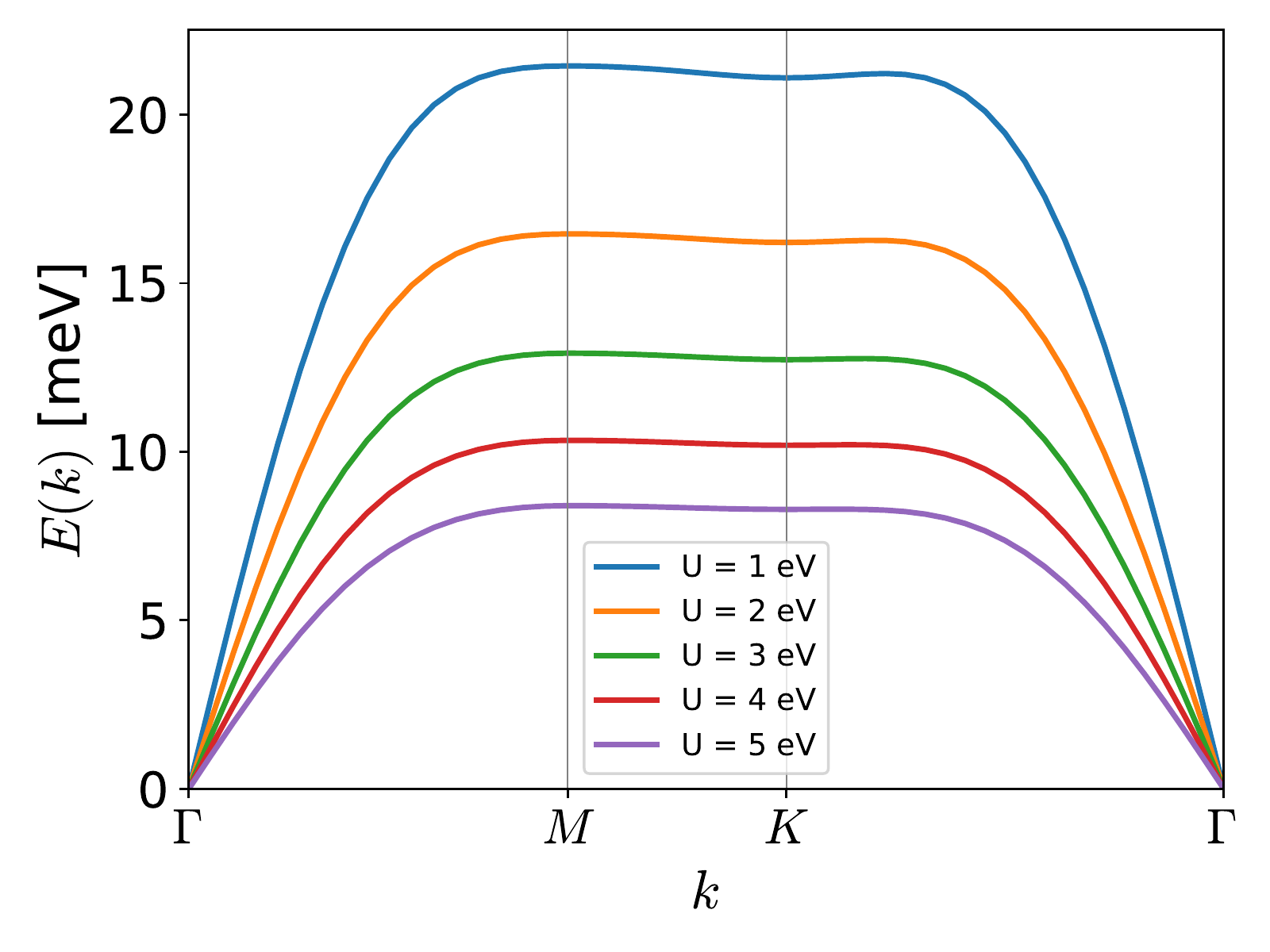}
	\caption{Magnon band structure of MnPS$_3$ using exchange coupling constants obtained with different values of the Hubbard correction $U$. The energy was calculated from Eq. \eqref{eq:dispersion}.}
	\label{fig:bands}
\end{figure}
The values of exchange and anisotropy constants will have crucial influence on any magnetic property calculated for the system. As an example, Fig. \ref{fig:bands} shows the magnon dispersion relation of MnPS$_3$ calculated from the Heisenberg model \eqref{eq:H} using different values of U (see Appendix for details). The band width increases by a factor of 2.5 when the value of U is decreased from U = 5 eV to U = 1 eV. A band width of 12 meV has been determined from inelastic neutron scattering \cite{Wildes1998} and seems to agree well with the calculated dispersion relation using U = 3 eV. This is of course expected since the exchange parameters are in agreement with the experimental ones that were extracted from the measured magnon dispersion.
\begin{figure}[t]
	\includegraphics[width = 0.45\textwidth]{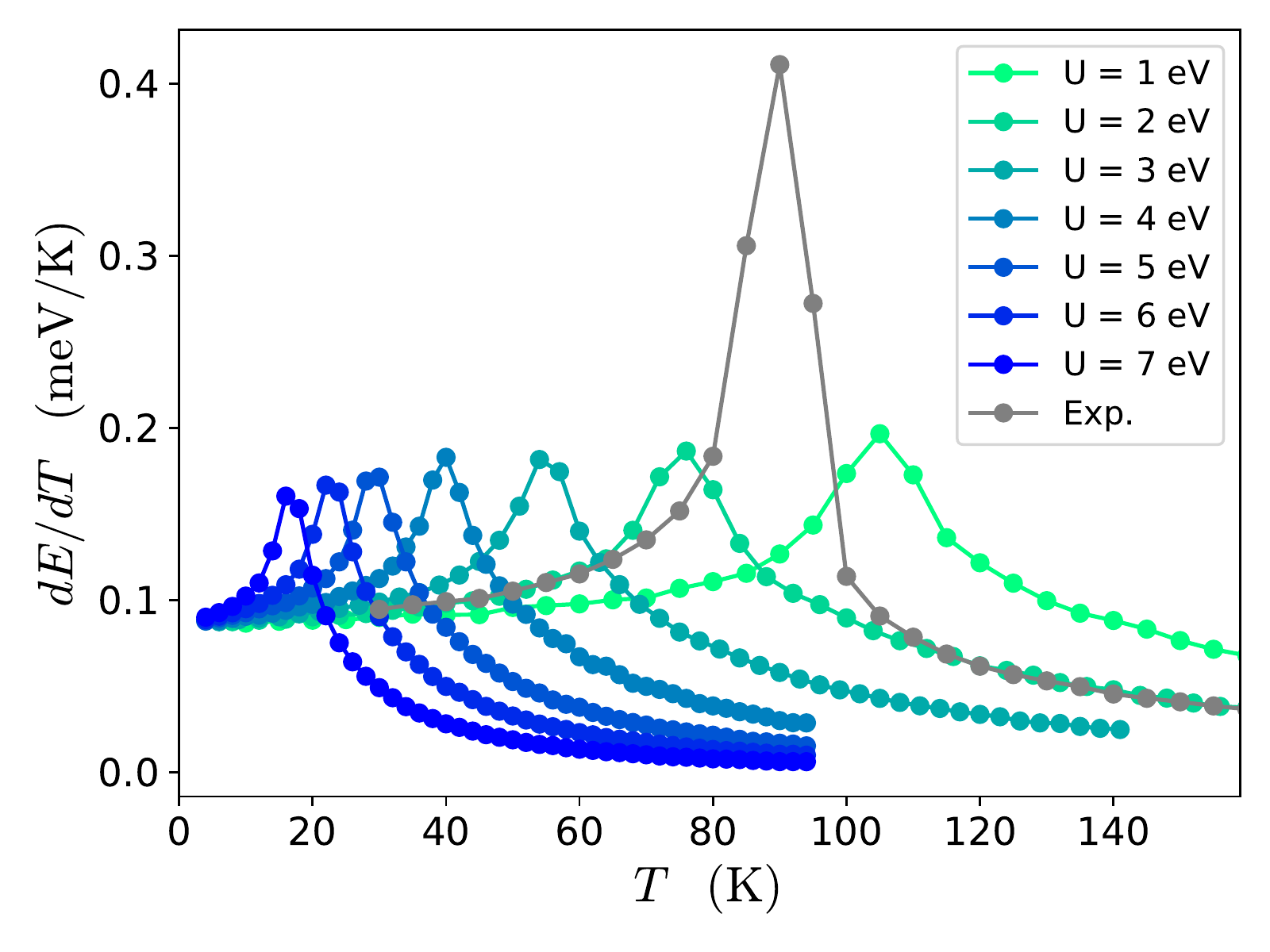}
	\caption{Heat capacity of FePS$_3$ evaluated from classical Monte Carlo simulations with Heisenberg parameters extracted from DFT with different values of U.}
	\label{fig:heat_capacity}
\end{figure}

In contrast to the exchange parameters, the predicted single-ion anisotropy differs from experimental values by more than an order magnitude for FePS$_3$ and NiPS$_3$. Experimentally the value is determined from the spin-wave gap of the bulk material, which is assumed to originate from single-ion anisotropy. The single-ion anisotropy parameters have thus been estimated to 2.7, 0.3 and 0.009 meV for FePS$_3$, NiPS$_3$, and MnPS$_3$ respectively \cite{Lancon2018}. 
The theoretical predictions appear to be much to small irrespective of the value chosen for the Hubbard parameter. Moreover, only the case of FePS$_3$ yields a theoretical prediction of a positive value of the single-ion anisotropy corresponding to an easy axis orthogonal to the atomic plane. In contrast, experiments predict all three bulk materials have positive values. Due to the small magnitude of the interlayer exchange coupling constants \cite{Lancon2018} it does not seem likely that this discrepancy originates from the fact that the experimental values refer to the bulk materials. There may be other effects contributing to the spin-wave gap that are not accounted for in the fit to spin-wave spectra, but it is far from clear how such effects could give rise to an order of magnitude larger spin-wave gaps compared to the experimentally determined values. On the other hand, spin-orbit effects are usually well accounted for in DFT and is not highly sensitive to the choice of functional so it is not obvious why DFT would make an order of magnitude error for these materials either. We have tested that the single-ion anisotropy does not change significantly when using experimental lattice parameters instead of relaxed structure (we get A=0.116 meV for FePS$_3$ using the experimental alttice constant and U=2 eV). For now the origin of this discrepancy remains an open question.

The magnetic anisotropy plays a crucial role in the magnetic order for 2D materials. In particular, an easy axis is required for a 2D material to exhibit magnetic order at finite temperatures. However, the critical temperature has a logarithmic dependence on the magnetic anisotropy and critical temperatures will thus not be very sensitive to the magnitude of the single-ion anisotropy constant. In Fig. \ref{fig:heat_capacity}  we show classical Monte Carlo simulations of the heat capacity of FePS$_3$ using the parameters obtained from DFT with different Hubbard corrections. The Heat capacity has a peak at the Néel temperature, which is seen to have a strong dependence on U. We also show a simulation where the experimental parameters were used. The main difference between this set of parameters and the calculated ones with U=2 eV is the single-ion anisotropy, which is more than 20 times larger as determined from experiments. The predicted critical temperature, however, is only slightly larger compared to the theoretical results with U=2 eV, but the heat capacity is more sharply peaked due to the stronger "Ising-like" nature of the material resulting from the experimental parameters. 

\begin{figure}[t]
	\includegraphics[width = 0.45\textwidth]{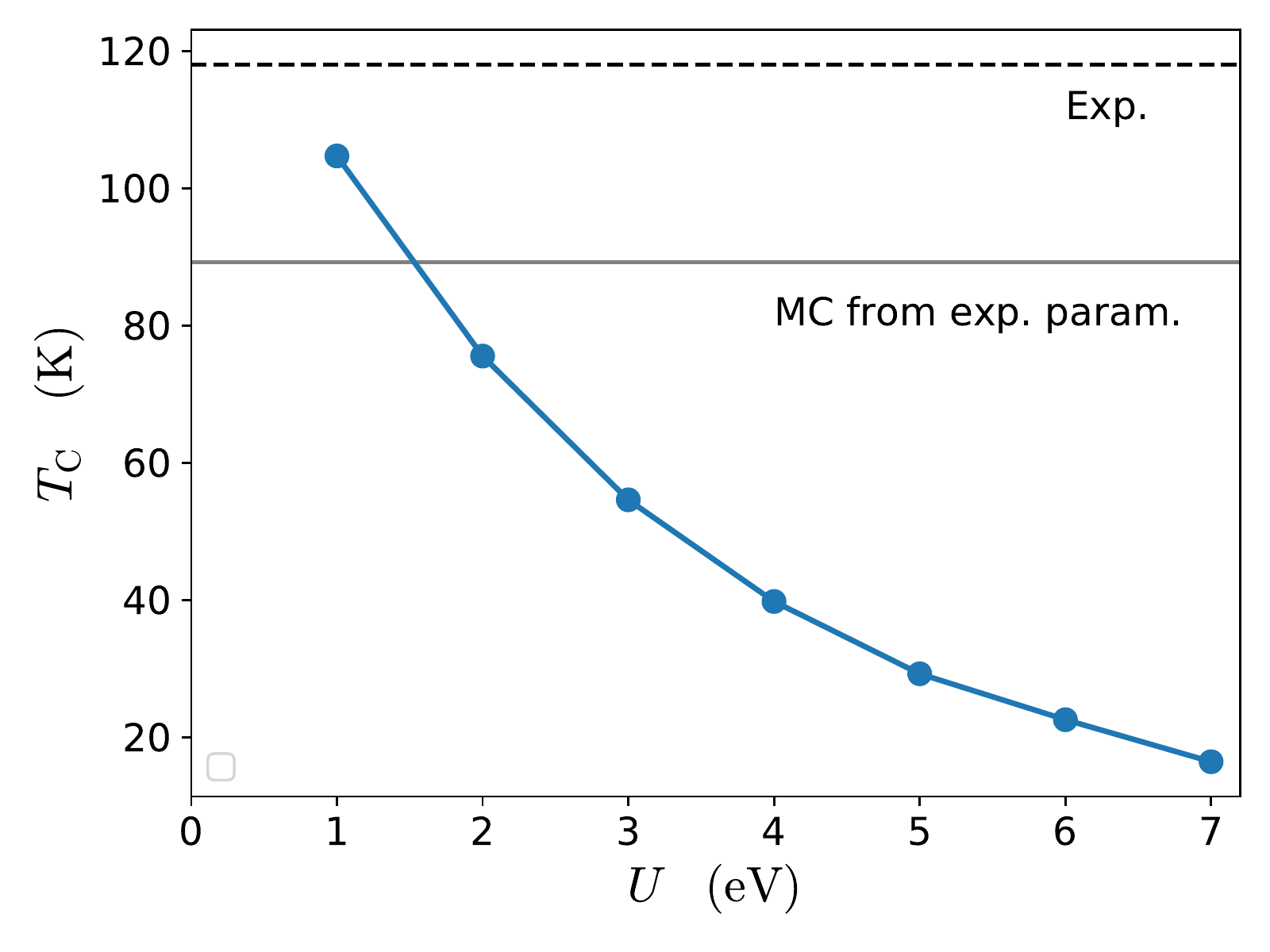}
	\caption{Critical temperature for FePS$_3$ evaluated from Monte Carlo simulation with Heisenberg parameters extracted from DFT+U using different values of U. The experimentally determine critical temperature for monolayer FePS$_3$ \cite{Lee2016} is shown by the dashed line.}
	\label{fig:Tc}
\end{figure}

In Fig. \ref{fig:Tc} we have extracted the critical temperatures from the Monte Carlo simulations of the heat capacity, which are plotted as a function of U. First of all we note that the critical temperature obtained from the experimental parameters gives 89 K, which is somewhat lower than the experimentally determined value of 118 K. From DFT it appears that simulation with $U<1$ is required to reproduce the experimental critical temperature, but this could be due to a strong underestimation of the single-ion anisotropy. We stress again that theoretical predictions are strongly dependent on the chosen value for the Hubbard correction.

\section{Discussion}\label{sec:discussion}
We have presented the Heisenberg parameters of FePS$_3$, NiPS$_3$, and MnPS$_3$ as predicted by DFT using the PBE+U approach. It was demonstrated that the magnitude of the parameters depend crucially on the chosen value of U. This is important to bear in mind since the value of U is often chosen to reproduce a particular experimental signature of the material. For example, in Ref. \cite{Qiu2021} a value of U=0.5 eV was chosen in in order to reproduce STS spectra, whereas values 2-4 eV are more common for accurate extraction of structural properties \cite{Sivadas2018}. It is however, reassuring that the experimental exchange constants are well reproduced by a particular value of U, which is in the range of commonly applied values. 

The observable consequences resulting from different Hubbard corrections were exemplified by the spin-wave band-width of MnPS$_3$ and the Néel temperature of FeP$_3$. The experimental band width of MnP$_3$ was trivially reproduced by U=3 eV calculations since the experimental parameters were extracted from the dispersion and U=3 eV yields parameters in good agreement with the experimental ones. The calculated Néel temperature of FePS$_3$ exhibits a strong dependence of U, which is naturally inherited from the exchange coupling constants. However, in order to reproduce the experimental Néel temperature a value of U$\sim$0.8 eV is required, which yields exchange constants that are not in agreement with the experimental values. In fact, the experimental parameters themselves yields a Néel temperature that is 15 {\%} smaller than the experimental one. The simplest explanation could be that the bulk material simply has weaker intralayer exchange constants than the 2D material. Such effects should be straightforward to unravel with DFT but is beyond the scope of the present work. Another possibility is the existence of important Heisenberg terms such as biquadratic exchange and dipolar interactions that are simply not accounted for in the present model \cite{Hoffmann2020}. Finally, it is possible that classical Monte Carlo simulations are insufficient to describe the Néel temperature accurately.

It is highly disturbing that the calculated anisotropy constants deviate from experimental values by more than an order of magnitude. We note, however, that the experimental parameters were derived from the spin-wave gap, which for anti-ferromagnets depends on the exchange constants as well as the anisotropy parameters \cite{Torelli2020}. Thus if additional Heisenberg terms are present they are expected to modify the spin-wave gap, which would yield different predictions for the experimental single-ion anisotropy parameters. In addition, many-body effects have recently been shown to play a crucial role for the gap opening between the acoustic and optical branches in CrI$_3$ \cite{Ke2021} - even without spinorbit coupling. Although, the spin-wave gap has to vanish in the absence of spinorbit coupling it is not unlikely that similar effects could play an important role in determining the {\it size} of the gap (and thus the predicted anisotropy constants) once spinorbit coupling is introduced. We leave these open questions to future work.

\section*{Appendix}
Here we present the non-interacting magnon spectrum of the Néel state on the Honeycomb lattice with three exchange coupling constants. The Heisenberg Hamiltonian is written as
\begin{align}\label{H}
H=&-J_1\sum_{\langle ij\rangle}\mathbf{S}_{ai}\cdot\mathbf{S}_{bj}-\frac{J_2}{2}\sum_{\langle\langle ij\rangle\rangle}\mathbf{S}_{ai}\cdot\mathbf{S}_{aj}\notag\\
&-\frac{J_2}{2}\sum_{\langle\langle ij\rangle\rangle}\mathbf{S}_{bi}\cdot\mathbf{S}_{bj}-J_3\sum_{\langle\langle\langle ij\rangle\rangle\rangle}\mathbf{S}_{ai}\cdot\mathbf{S}_{bj},
\end{align}
where $a$ and $b$ denotes the two in-equivalent sites in the unit cell. Performing the usual Holstein-Primakoff transformation to second order in raising and lowering operators yield
\begin{align}
H =& E_0 + H_0\\
E_0=&NS^2(N_1J_1-N_2J_2+N_3J_3)\\
H_0=&-SJ_1\sum_{\langle ij\rangle}\Big(a_ib_j+a_{i}^\dag b_j^\dag+a_i^\dag a_i+b_j^\dag b_j\Big)\notag\\
&-SJ_2\sum_{\langle\langle ij\rangle\rangle}\Big(a_i^\dag a_j+b_i^\dag b_j-a_i^\dag a_i-b_i^\dag b_i\Big)\notag\\
&-SJ_3\sum_{\langle\langle\langle ij\rangle\rangle\rangle}\Big(a_ib_j+a_{i}^\dag b_j^\dag+a_i^\dag a_i+b_j^\dag b_j\Big),
\end{align}
where $N$ is the number of unit cells and $N_n$ is the number of $n$'th nearest neighbors.
%where $N$ is the total number of unit cells and we defined
%\begin{align}
%J({\mathbf{q}})=\sum_{n=1}^\infty J_n\sum_{\mathbf{R}_n}e^{i\mathbf{q}\cdot\mathbf{R}_n},
%\end{align}
%where $J_n$ is the $n$'th nearest neighbor exchange coupling and $\mathbf{R}_n$ are the lattice vectors connecting a site to its $n$'th nearest neighbors. Thus interactions connecting sites in unit cells indexed by $i,j$ can be written as $\mathcal{J}_{abij}=\mathcal{J}_{ab}(\mathbf{R}_n)$ where $R_n$ is the lattice vector connecting the unit cells containg site $ai$ and $bj$.
We emphasize that the Néel state is not an eigenstate of the Heisenberg Hamiltonian but its expectation value is given by $E_0$, which coincides with the classical minimum energy. 

We then introduce the Fourier transforms
\begin{align}
a_{i}&=\sqrt{\frac{1}{N}}\sum_{\mathbf{q}}e^{i\mathbf{q}\cdot\mathbf{R}_i}a_\mathbf{q}\label{a_q}\\
b_{j}&=\sqrt{\frac{1}{N}}\sum_{\mathbf{q}}e^{-i\mathbf{q}\cdot\mathbf{R}_j}b_\mathbf{q}\label{b_q},
\end{align}
where $\mathbf{R}_i$ are the positions of sublattice a sites and $\mathbf{R}_j$ are the positions of sublattice b sites.
Inserting into $H_0$ yields
\begin{align}
H_0=&-SJ_1\sum_{\mathbf{q}\Delta_1}\Big(e^{-i\mathbf{q}\cdot\mathbf{R}_{\Delta_1}}a_{\mathbf{q}}b_{\mathbf{q}}+e^{i\mathbf{q}\cdot\mathbf{R}_{\Delta_1}}a_{\mathbf{q}}^\dag b_{\mathbf{q}}^\dag\Big)\notag\\
&-SN_1J_1\sum_{\mathbf{q}}\Big(a_{\mathbf{q}}^\dag a_{\mathbf{q}}+b_{\mathbf{q}}^\dag b_{\mathbf{q}}\Big)\notag\\
&-SJ_2\sum_{\mathbf{q}\Delta_2}\Big(e^{i\mathbf{q}\cdot\mathbf{R}_{\Delta_2}}a^\dag_{\mathbf{q}}a_{\mathbf{q}}+e^{-i\mathbf{q}\cdot\mathbf{R}_{\Delta_2}}b_{\mathbf{q}}^\dag b_{\mathbf{q}}\Big)\notag\\
&+SN_2J_2\sum_{\mathbf{q}}\Big(a_{\mathbf{q}}^\dag a_{\mathbf{q}}+b_{\mathbf{q}}^\dag b_{\mathbf{q}}\Big)\notag\\
&-SJ_3\sum_{\mathbf{q}\Delta_3}\Big(e^{-i\mathbf{q}\cdot\mathbf{R}_{\Delta_3}}a_{\mathbf{q}}b_{\mathbf{q}}+e^{i\mathbf{q}\cdot\mathbf{R}_{\Delta_3}}a_{\mathbf{q}}^\dag b_{\mathbf{q}}^\dag\Big)\notag\\
&-SN_3J_3    \sum_{\mathbf{q}}\Big(a_{\mathbf{q}}^\dag a_{\mathbf{q}}+b_{\mathbf{q}}^\dag b_{\mathbf{q}}\Big)\\
=&-S\sum_{\mathbf{q}}\Big(N_1J_1+N_2J_2[\gamma_2(\mathbf{q})-1]+N_3J_3\Big)\notag\\
&\qquad\times\bigg[a_{\mathbf{q}}^\dag a_{\mathbf{q}}+b_{\mathbf{q}}^\dag b_{\mathbf{q}}+\tilde\gamma_\mathbf{q}a_{\mathbf{q}}b_{\mathbf{q}}+\tilde\gamma_\mathbf{q}^*a_{\mathbf{q}}^\dag b_{\mathbf{q}}^\dag\bigg]
\end{align}
where
\begin{align}
&\tilde\gamma(\mathbf{q})=\frac{N_1J_1\gamma_1(\mathbf{q})+N_3J_3\gamma_3(\mathbf{q})}{N_1J_1+N_2J_2[\gamma_2(\mathbf{q})-1]+N_3J_3},\\
&\gamma_n(\mathbf{q})=\frac{1}{N_{n}}\sum_{\Delta_n}e^{-i\mathbf{q}\cdot\mathbf{R}_{\Delta_n}},
\end{align}
and $\mathbf{R}_{\Delta_n}$ are the vectors connecting the $n$'th nearest neighbor atoms. 

The Hamiltonian can now be diagonalized by the Bogliubov transformation
\begin{align}
a_\mathbf{q}&=\cosh\theta_\mathbf{q}\alpha_\mathbf{q}-\sinh\theta_\mathbf{q}\beta_\mathbf{q}^\dag\label{eq:bogoliubov1}\\
b_\mathbf{q}&=-\sinh\theta_\mathbf{q}\alpha_\mathbf{q}^\dag+\cosh\theta_\mathbf{q}\beta_\mathbf{q}\label{eq:bogoliubov2}
\end{align}
where $\alpha_\mathbf{q}$ and $\beta_\mathbf{q}$ satisfy the usual bosonic commutator relations and $\tanh2\theta_\mathbf{q}=|\tilde\gamma(\mathbf{q})|$. The non-interacting part of the Hamiltonian then becomes
\begin{align}
H_0=S&\sum_{\mathbf{q}}\Big(N_1J_1+N_2J_2[\gamma_2(\mathbf{q})-1]+N_3J_3\Big)\\
&\times\bigg\{1-\sqrt{1-|\tilde\gamma(\mathbf{q})|^2}\Big(\alpha_{\mathbf{q}}^\dag \alpha_{\mathbf{q}}+\frac{1}{2}+\beta_{\mathbf{q}}^\dag \beta_{\mathbf{q}}+\frac{1}{2}\Big)\bigg\}.\notag
\end{align}
The new operators $\alpha_{\mathbf{q}}$ and $\beta_{\mathbf{q}}$ define a new "Non-interacting magnon" ground state defined by $\alpha_\mathbf{q}|0\rangle_{\mathrm{NIM}}=\beta_\mathbf{q}|0\rangle_{\mathrm{NIM}}=0$. This state has a lower energy than the Néel state and it is given by
\begin{align}
E_0^{\mathrm{NIM}}=E_0 + SN\bigg\langle\Big(N_1J_1+N_2J_2[\gamma_2(\mathbf{q})-1]+N_3J_3\Big)\notag\\
\times\Big(1-\sqrt{1-|\gamma(\mathbf{q})|^2}\Big)\bigg\rangle_{BZ}.
\end{align}
We have written the sum as a BZ average denoted by $\langle\ldots\rangle_{BZ}$ and multiplied by $N$ since the $\mathbf{q}$-sum contains $N$ terms. 

Finally, the single magnon excited states have an energy {\it relative} to the NIM state given by
\begin{align}\label{eq:dispersion}
\varepsilon_\mathbf{q}=-S\Big(N_1J_1+N_2J_2[\gamma_2(\mathbf{q})-1]+N_3J_3\Big)\sqrt{1-|\gamma(\mathbf{q})|^2}.
\end{align}

\bibliography{references}

\end{document}